\documentclass[11pt]{article} 

\usepackage[utf8]{inputenc} 

\usepackage{geometry} 
\usepackage{graphicx} 

\usepackage{booktabs} 
\usepackage{array} 
\usepackage{paralist} 
\usepackage{verbatim} 

\usepackage{fancyhdr} 
\usepackage{sectsty}
\allsectionsfont{\sffamily\mdseries\upshape} 

\usepackage[nottoc,notlof,notlot]{tocbibind} 
\usepackage[titles,subfigure]{tocloft} 

\usepackage{multirow,array}
\usepackage{float} 
\usepackage{amsmath}
\usepackage{amsfonts}
\usepackage{amsthm}

\let\origref \ref
\def \ref#1{\textbf{\origref{#1}}}

\usepackage{subcaption}

\newtheoremstyle{break}
  {\topsep}{\topsep}%
  {\itshape}{}%
  {\bfseries}{}%
  {\newline}{}%
\theoremstyle{break}

\usepackage[natbibapa]{apacite}
\setcitestyle{authoryear,open={(},close={)}}

\usepackage{comment}
\usepackage{xcolor}
\definecolor{darkorange}{rgb}{0.8, 0.4, 0.0}  

\usepackage{graphicx,multirow}

\usepackage[title,toc,titletoc]{appendix}

\usepackage{tikz}

\title{Effective Theory Building and Manifold Learning}
\author{David Peter Wallis Freeborn}

\begin{document}
\maketitle

\begin{abstract}
Manifold learning and effective model building are generally viewed as fundamentally different types of procedure. After all, in one we build a simplified model of the data, in the other, we construct a simplified model of the another model. Nonetheless, I argue that certain kinds of high-dimensional effective model building, and effective field theory construction in quantum field theory, can be viewed as special cases of manifold learning. I argue that this helps to shed light on all of these techniques. First, it suggests that the effective model building procedure depends upon a certain kind of algorithmic compressibility requirement. All three approaches assume that real-world systems exhibit certain redundancies, due to regularities. The use of these regularities to build simplified models is essential for scientific progress in many different domains.
\end{abstract}

\section{Introduction}
\label{sec:introduction}

Manifold learning is a very widespread family of dimensional reduction techniques in machine learning, in which high-dimensional data is projected onto a lower-dimensional manifold, while preserving some salient properties of the original data \citep{hinton2002stochastic, maaten2008visualizing, tenenbaum2000global, hinton2006reducing, mcinnes2018umap, roweis2000nonlinear, LLEpaper2001}. This technique is based on the assumption that many high-dimensional datasets contain regularities that allow them to be conveniently compressed or summarized with a simpler model. Likewise, effective theory or model construction is a family of techniques in physics and the computational sciences, in which a high-dimensional theory or model is reduced to a lower-dimensional one. Effective theory building is commonly used in quantum field theory, where the mathematical problems have led to the construction of lower-dimensional effective field theories, and in many computational sciences, where there are many high-dimensional models, highly insensitive to the vast majority of parameter combinations \citep{Burgess2020, Duncan, Sloppy_dyson, Sloppy_RG, Sloppy_RG_info}.

Manifold learning and effective model building are generally viewed as fundamentally different types of procedure. After all, in one we build a simplified model of the data, in the other, we construct a simplified model of the another model \citep{sloppy_simplicity, sloppy_recent, sloppy_kullback}. Indeed, they use the term \textbf{model} to mean two importantly different things.
\begin{itemize}
    \item \textbf{Machine Learning Models} are functions that map from high-dimensional input data to lower-dimensional outputs, such as classifications or predictions.
    \item \textbf{Scientific/Computational Models} are mathematical representations of physical systems, typically mapping from theoretical parameters to observable predictions.
\end{itemize}

Nonetheless, I argue that certain kinds of high-dimensional effective model building, and effective field theory construction in quantum field theory, can be viewed as special cases of dimensional reduction techniques akin to manifold learning. I argue that this helps to shed light on some underlying principles shared by all of these techniques. First, it suggests that the effective model building procedure depends upon a certain kind of algorithmic compressibility requirement. All three approaches assume that real-world systems exhibit certain redundancies, due to regularities. The use of these regularities to build simplified models is essential for scientific progress in many different domains.

These topics have generated significant philosophical interest in recent years. There has been an ongoing debate over how effective theories and related methods can inform and refine scientific realism, particularly in the context of quantum field theory. Proponents of \emph{effective realism} \citep{Wallace_naivete, PorterWilliams, Fraser_Realism, Fraser_RealProblem, Fraser_Towards, Miller} argue that these methods can inform and refine a localized, theory-specific approach to realism by identifying the elements of quantum field theory (QFT) models that are empirically robust and likely to persist through scientific progress. However, this defense has been challenged by critics like \citet{Ruetsche}, who argue that while effective realism engages directly with successful aspects of current physics, it fails to fully mitigate skeptical challenges. Ruetsche suggests that these issues merely retreat to a different level rather than being resolved. Similarly, \citet{Sebastien_Rivat} contends that effective theories rely on intrinsic empirical limitations and infinite idealizations that constrain their scope to offer reliable ontological commitments. He argues that these idealizations, while useful for making accurate predictions within certain domains, pose significant challenges for ensuring the stability and approximate truth of theoretical representations through future theory changes.

Likewise, philosophers have debated the the related topic, reduction and emergence in the context of renormalization group methods (see section 9). \citet{Batterman2002, Batterman2011} argues that phenomena such as critical behavior and phase transitions require explanations that transcend simple deductive reductions, emphasizing the importance of renormalization group theory in understanding how macroscopic properties emerge from microscopic interactions. He contends that the renormalization group theory reveals how different scales interact and influence each other, demonstrating that certain macroscopic behaviors cannot be fully reduced to microscopic laws. Similarly, \citet{Morrison2012} highlights how renormalization group theory exemplifies the interplay between reduction and emergence in practice. Conversely, \citet{Butterfield_Intro} proposes that reduction and emergence are not mutually exclusive, arguing that these techniques provide a means to connect micro and macro levels, thereby reconciling reductionism with emergent properties.

In section \ref{sec:dimensional}, I explain the dimensional reduction. I introduce manifold learning as a particular case of this in section \ref{sec:manifold-learning}. In section \ref{sec:manifold-hypothesis}, I present the manifold hypothesis, and suggest one way to explicate it in a partly formal way. In section \ref{sec:sloppy}, I introduce the sloppy models program. In section \ref{sec:MBAM}, I argue that an effective model building technique, the manifold boundary approximation method can be viewed as akin to a special kind of manifold learning. In section \ref{sec:renormalization}, I introduce effective field theories, and in section \ref{sec:renormalizability}, I argue that it can be related to both the sloppy models program and manifold learning. I conclude by drawing some overall analogies between these approaches.

\section{Machine Learning and Dimensional Reduction}
\label{sec:dimensional}

Imagine that a machine learning specialist wants to build an artificial intelligence tool for recognizing handwritten numerical digits. As input data, they train their tool on the MNIST (Modified National Institute of Standards and Technology) database, a large collection of handwritten digits commonly used for training various image processing systems. The training data contains 60,000 $28 \times 28$ pixel images of handwritten digits ranging from 0 to 9 \citep{MNISTpaper, MNISTdatabase}. The aim is build a tool that can, in some sense, latch onto and generalize from key features of these handwritten digits, and which can then be applied to correctly interpret new handwritten images of digits, from outside of the training data.

In effect, the artificial intelligence tool serves as a \emph{model} of the data. We can think of such a model as a function, $f$, from a real-valued vector of the 784 pixels in each image, to a vector of ten real-valued output classifications, giving some measure of how likely the model thinks it is that the image represents each possible digit 0-9,\footnote{For instance, in a Bayesian model, these real-valued output classifications could represent probabilities.}

\begin{equation}
    f : \mathbb {R}^{784} \rightarrow \mathbb {R}^{10}.
\end{equation}

\noindent This general task, finding a function, mapping a real, $N$-dimensional data-vector to an $M<N$-dimensional output vector, is very common across machine learning. Indeed, almost any machine learning task can be represented as the task of finding a function of this form.\footnote{For example, we can represent almost any predictive AI task (e.g. image classification, speech recognition, natural language processing tasks such as sentiment analysis and machine translation, recommender systems, medical diagnosis, financial forecasting, etc.) \emph{or} generative AI task (e.g. text generation with a large language model or image generation with an adversarial network), as the task of finding a function of this form \citep{bishop2006pattern, hastie2009elements, lecun2015deep, vapnik1995nature, murphy2012machine, goodfellow2016deep}.} This is closely related to standard ways to think about model-building across the computational empirical sciences more generally.\footnote{For instance, see \citet{Breiman2001StatisticalMT}, or for related examples, see \citet{Sloppy_ubiquitous}, and see \citet{Williamson2009, Sozou2017, Sullivan2022} for some philosophical considerations.}

Our machine learning specialist might not merely seek the most predictively accurate model; often they will also want the model to be simple. Simpler models usually make lower demands on computational resources for training, inference and application; the results may be more robust to small modifications; they may be easier to interpret or explain; and they may be less inclined to overfit the training data, allowing for better generalizability to new data. Furthermore, a variety of technical problems are known to arise when the dimensionality of the data is very high compared to the number of datapoints, resulting in the so-called \emph{curse of dimensionality} \citep{bellman1957dynamic, bellman1961adaptive}.\footnote{Loosely speaking, the curse of dimensionality problems refer to the general observation that, as the dimensionality of the data grows, the volume grows so rapidly that finite data becomes sparsely distributed and increasingly orthogonal, making distance measures less able to extract useful information.}

Fortunately, the key features higher-dimensional real-world data can often be conveniently summarized by models with lower numbers of parameters. For instance, the salient variations in the MNIST handwritten digits might be summarizable by a much smaller number of factors or dimensions - rather than specifying each individual pixel, perhaps we can summarize them with a smaller number of identifiable curves, loops and lines. This task is at the heart of machine learning, algorithmic compression, and computational model-building more generally.

Thus, an obvious approach to simplify the model would be to first build a  lower-dimensional model of the data. That is, instead of applying our model, $f:\mathbb{R}^N \rightarrow \mathbb{R}^M$, to the data directly, we could first reduce the dimensionality of the data with a model, $m : \mathbb{R}^N \rightarrow \mathbb{R}^K$, and then apply a simpler model,  $g : \mathbb{R}^K \rightarrow \mathbb{R}^M$, with $N < K < M$. If the two processes give the same outputs, then we can think of $f$ as the composition of $g$ and $m$, as in figure \ref{fig:category}. However, in reality this is an unrealistic assumption: the two processes should give \emph{almost} the same outputs, but some information will be lost when compressing the model. We call the high-dimensional space the \textbf{feature space} of the data, and the low-dimensional space the \textbf{latent space}. This process is now widespread in machine learning (see \citealt{pearson1901lines, fisher1936use, izenman1975reduced} for some historical background to these techniques, and for contemporary examples, see \citealt{jolliffe2016principal, hinton2002stochastic, maaten2008visualizing, tenenbaum2000global, hinton2006reducing, mcinnes2018umap, roweis2000nonlinear, LLEpaper2001}).

\begin{figure}[htp]
\centering
\begin{tikzpicture}
\node (RN) at (0,0) {$\mathbb{R}^N$};
\node (RK) at (6,0) {$\mathbb{R}^K$};
\node (RM) at (12,0) {$\mathbb{R}^M$};
\node at (0,1) {\textit{Feature space}};
\node at (0,0.5) {\textit{(data)}};
\node at (6,1) {\textit{Latent space}};
\node at (6,0.5) {\textit{(simplified data)}};
\node at (12,1) {\textit{Output space}};
\node at (12,0.5) {\textit{(model outputs)}};
\draw[->] (RN) -- (RK) node[midway,above] {$m$} node[midway,below] {\textit{Dimensional reduction}};
\draw[->] (RK) -- (RM) node[midway,above] {$g$} node[midway,below] {\textit{Simplified model}};
\draw[->,bend right] (RN) to node[midway,below] {\textit{Original model}} node[midway,below=0.5cm] {$f = g \circ m$} (RM);
\end{tikzpicture}
\caption{A category theoretic representation of the direct and simplified modeling approaches, assuming that they give the same outputs. Here, $\mathbb{R}^N$ is the feature space, $\mathbb{R}^K$ gives a latent space offering a simplified representation of the data, and $\mathbb{R}^M$ is the output space. The functions $f$, $m$, and $g$ represent the original predictive model, the dimensionality-reducing model, and the simplified predictive model, respectively.}
\label{fig:category}
\end{figure}

The key is that our dimensional reduction model, $m$, must preserve certain salient local or global features of the data, even as it throws out some of the information contained in the original data. The salient features encoded in the data might vary, depending on the task at hand. They might include  geometric properties (such as distances between data points, angles or local curvatures) or topological properties (including shape and connectivity features like clusters, holes, and loops). For instance, with our MNIST data, perhaps different ways of writing the same digit (like a closed `4' versus an open `4'.) might form distinct subclusters within a larger cluster. Topological information could help in understanding the transition between different writing styles (for example, a curly `9' might continuously morph into a straight `9'). We define cost functions to measure how well certain salient properties of the data are preserved by the function $m$.

\section{Manifold Learning}
\label{sec:manifold-learning}

Roughly speaking, a manifold is a topological  space that locally resembles flat Euclidean space.\footnote{More fully, an $n$-dimensional topological manifold is a topological space $\mathcal{M}$ which satisfies three conditions: First, it must be locally Euclidean, meaning that for every point $p$ in $\mathcal{M}$, there exists an open neighborhood $U$ around $p$ that is homeomorphic to an open subset of Euclidean space $\mathbb{R}^n$, where $n$ is a fixed integer representing the manifold's dimension. This ensures that sufficiently small neighborhoods in $\mathcal{M}$ locally resemble flat Euclidean space. Second, $\mathcal{M}$ must obey the Hausdorff condition, that for any two distinct points in $\mathcal{M}$, there exist disjoint open neighborhoods. This ensures that points can always be separated by open sets. Finally, $\mathcal{M}$ must be second-countable, meaning it possesses a countable basis for its topology. See \citet{guillemin1974, hirsch1994} for further details.
} Computer scientists have found that real-world data in $\mathbb{R}^N$ often lie close to a lower-dimensional manifold, $\mathcal{M}$, which can be embedded into $\mathbb{R}^K$, $K<N$. A suitable embedding function from $\mathbb{R}^N$ to $\mathbb{R}^K$ could provide a very convenient model of the data, one that preserves salient topological properties also being sensitive to nonlinear relationships between datapoints. In practice, manifolds seem to carry just the right amount of structure for this task \citep{hinton2002stochastic, maaten2008visualizing, tenenbaum2000global, hinton2006reducing, mcinnes2018umap, roweis2000nonlinear, LLEpaper2001}.

Roughly speaking, \textbf{manifold learning} is a family of dimensional reduction algorithms that progress according to the following scheme. 
\begin{itemize}
    \item We begin with the dataset with datapoints ${x_i} \in \mathbb{R}^N$. We posit that there exists a manifold $\mathcal{M}$ of dimension $K < N$, embedded in $\mathbb{R}^N$, such that the data points lie on or close to the manifold.
    \item The goal is to find an embedding function $m : \mathbb{R}^N \rightarrow \mathbb{R}^K$  that projects each high-dimensional datapoint in the feature space onto the $K$-dimensional latent space, $\mathbb{R}^K$.\footnote{The function $m$ is an embedding if it is a smooth, injective, immersion, whose underlying continuous function is a homeomorphism onto its image (see \citealt[pages 21-29]{hirsch1994} for further details).} It maps $\mathbb{R}^N$ onto $\mathbb{R}^K$ such that the images $m(x_i)$ preserve the intrinsic geometric and topological structure of the original data $x_i$ on the manifold $\mathcal{M}$, within the constraints of the reduced dimensionality.
    \item We define a cost function,
    \begin{equation}
    C : \mathbb{R}^N \times \mathbb{R}^K \rightarrow \mathbb{R},
    \end{equation}
    \noindent which assigns a real number to each pair of points, one from the feature space and one from the latent space, designed to measure how well a map preserves salient geometric and topological features of the data (i.e. structural features of $\mathcal{M}$).
    \item We find an embedding, $m$, that minimizes the cost function.
    \item Finally, the reduced-dimension data points, $y_i$, are represented in the lower-dimensional latent space  by their images under the embedding, $m(x_i) = y_i \in \mathbb{R}^K$.
\end{itemize}

When applying this procedure, it is essential to avoid \textbf{overfitting}, in which the model captures noise in the data, thereby failing to provide generalizable insights about the data. In the extreme case, without any procedures to avoid overfitting, we might represent all the data with a one-dimensional manifold, a curve passing through each datapoint. While this curve would perfectly `fit' the data, it would fail to capture the simpler, underlying structures that we seek to learn

Therefore any manifold learning technique will generally require us to implement some techniques to prevent overfitting, often in the form of a smoothness constraint. There are three widely-used (non-exclusive) approaches to this.

\begin{itemize}
    \item \textbf{Constraints on the manifold:} We explicitly restrict the class of allowable manifolds to those meeting certain smoothness criteria. I will discuss one example, the \emph{reach constraint} in section \ref{sec:manifold-hypothesis}.\footnote{These constraints are important in the theoretical studies of manifold learning. But they are not so widely used in practically useful algorithms, as directly enforcing such constraints can be computationally expensive \citep{Belkin2006, fefferman2016manifold, manifold-reach}.}
    \item \textbf{Cost Function:} We favour smoother and simpler manifolds implicitly in the cost function. For example, in the \emph{Locally Linear Embedding} algorithm \citep{roweis2000nonlinear}, we mitigate the effect of noise by approximating each point as a linear combination of its nearest neighbors. These nearest neighbours are likely to be part of the same smooth patch of the manifold. Overly contorted manifolds are often disfavored by this process.
    \item \textbf{Regularization:} We further modify the cost function to penalize insufficiently smooth solutions. Such a cost function might look like, $C_\text{total}(x_i, y_i) = C_\text{base}(x_i, y_i) + R(y_i)$, Where $C_\text{base}$ measures how well the low-dimensional representation $y_i$ preserves the structure of the original data $x_i$, and $R$ is a regularization term that increases with the 'roughness' of the embedding \citep{hastie2009elements}. 
\end{itemize}

\begin{figure}[htp]
    \centering
    \begin{subfigure}{0.7\textwidth}
        \includegraphics[width=\linewidth]{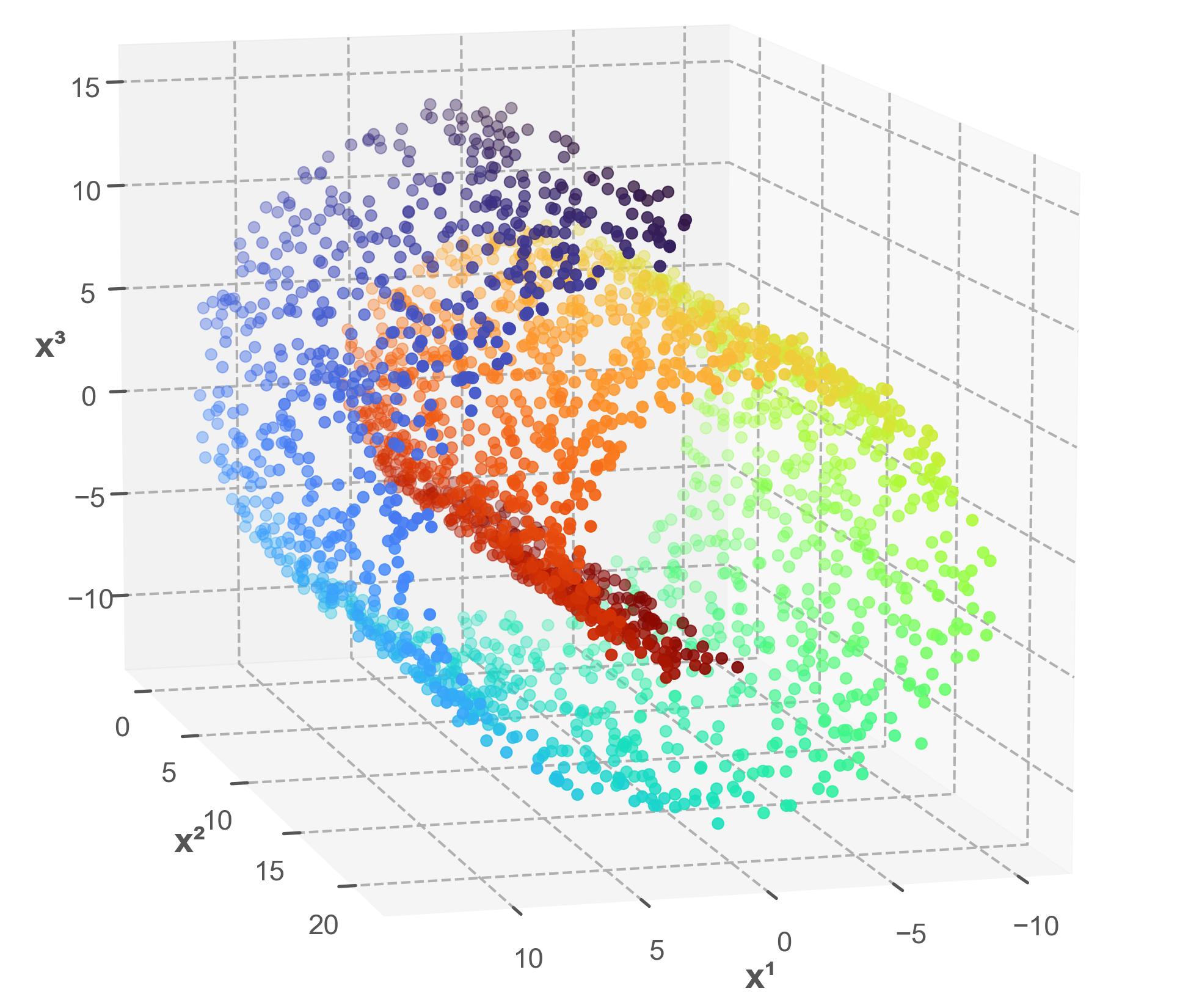}
        \caption{2500 randomly generated datapoints in $\mathbb{R}^3$, lying close to the swiss-roll surface. The colours are for visualization only.}
        \label{fig:swissroll}
    \end{subfigure}
    
    \begin{subfigure}{0.7\textwidth}
        \includegraphics[width=\linewidth]{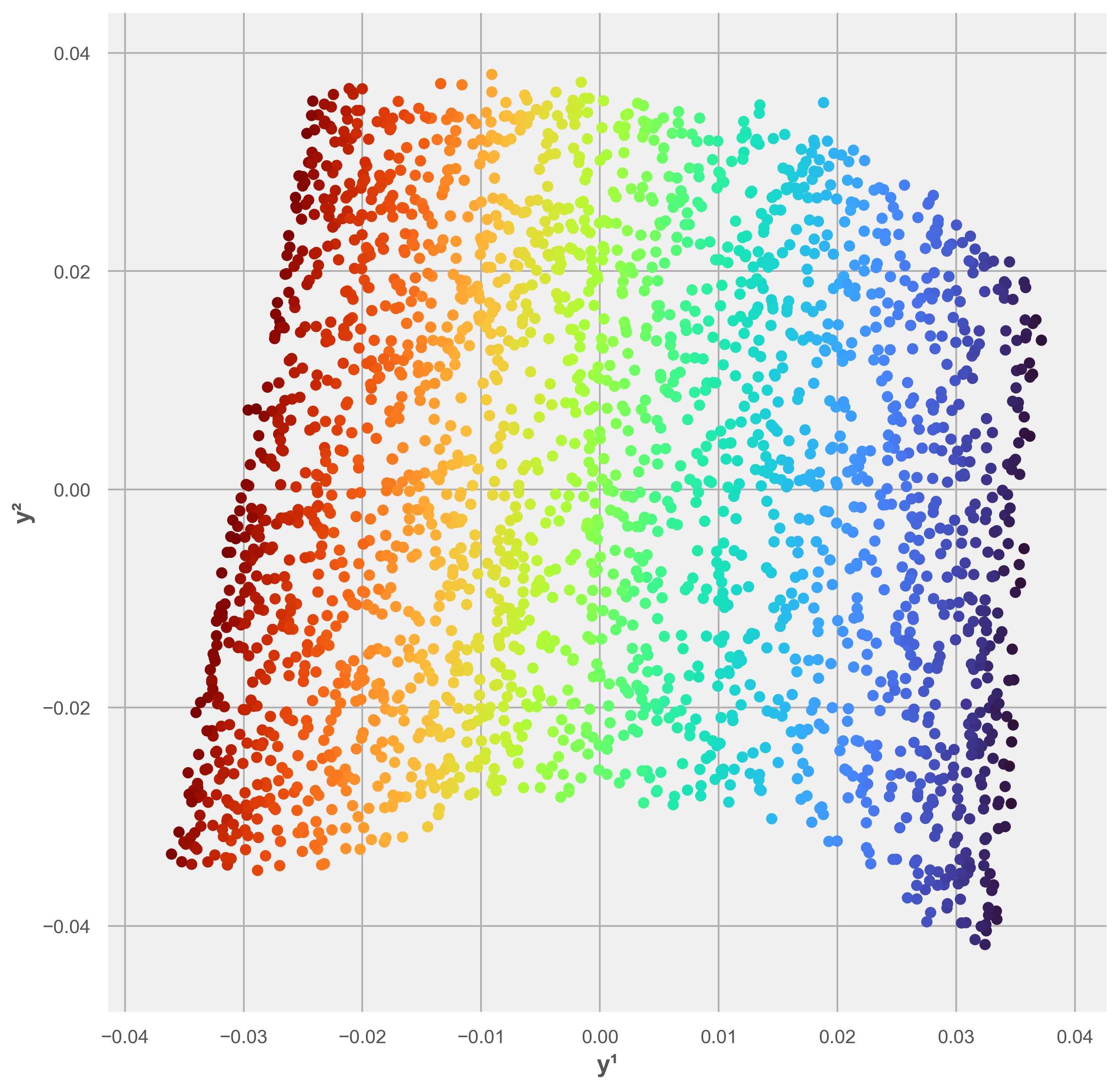}
        \caption{The datapoints transformed under the LLE ($k=20$) algorithm, represented in $\mathbb{R}^2$. Corresponding datapoints keep their colour from figure \ref{fig:swissroll}.}
        \label{fig:swisspancake}
    \end{subfigure}
    
    \label{fig:combined}
\end{figure}

Let us consider a very simple example of manifold learning (see \citealt{tenenbaum2000global}) and loosely show how to apply one possible local manifold-learning algorithm,\footnote{We can loosely distinguish two kinds of manifold learning algorithm, local and global methods \citep{cayton2005algorithms}. For local methods, the cost function considers the placement of each point with respect to its neighbors, whereas for global methods tend to consider the relative placement of all points.} \emph{Locally Linear Embedding} \citep{roweis2000nonlinear}. Suppose that our data is composed of points in a three-dimensional feature space, $\mathbb{R}^3$. Further suppose that the datapoints tend to lie close to a surface, described by the \emph{swiss roll} parametric equations,

\begin{align}
    x^1 &= t \cos (t) \\
    x^2 &= s \\
    x^3 &= t \sin (t),
\end{align}

\noindent where $x^{1,2,3}$ are some choice of the three coordinates, and $t$ and $s$ are parameters (see figure \ref{fig:swissroll}).\footnote{The data was generated using the \texttt{scikit-learn} dataset, \texttt{make-swiss-roll} \citep{scikit-learn}.}

Let us suppose that we want to reduce the dimensionality of this data to a latent space of just two dimensions, whilst trying to preserve the geometric features of the original global non-linear structure. If the datapoints lie near a two-dimensional manifold as we hope, then there should be a linear mapping from the coordinates of each neighbourhood to coordinates on the manifold which preserves this structure. So one approach could be to proceed as follows. First, we identify the $k$ nearest neighbors for each point in the dataset, and some choice of integer, $k$, using on Euclidean distance in the $\mathbb{R}^3$ space. We assume that each data point and its neighbors lie to close to a \emph{locally linear} patch of the manifold. Then, each point $x_i$ can be reconstructed from a linear sum of the coordinates of its neighbor, $x_js$. As such, we minimize the cost function, $\varepsilon$,

\begin{equation}
    \varepsilon = \lvert x_i - \sum_j W_{i,j} x_j \rvert ^2,
\end{equation}

\noindent where the weights $W_{i,j}$ give the contribution of the $j$th data point to the $i$th reconstructed point. 

Finally, we find the corresponding points, $y_i$, in the latent space, $\mathbb{R}^2$, that best preserve these local weights. This is done by minimizing another cost function, $\phi$,

\begin{equation}
    \phi = \lvert y_i - \sum_j W_{i,j} y_j \rvert ^2,
\end{equation}

\noindent where $y_i$, $y_j$ are the corresponding lower-dimensional embeddings of $x_i$ and $x_j$ respectively. The result is a lower-dimensional expression of the original data, preserving some of its original geometric features, albeit with some (hopefully small) loss of information. Figure \ref{fig:swisspancake} shows the application of this algorithm to the data from figure \ref{fig:swissroll}: our swiss roll has been unfurled and flattened into a pancake.

\section{The Manifold Hypothesis}
\label{sec:manifold-hypothesis}

It is widely posited that all such manifold learning techniques share a common \emph{fundamental assumption} \citep{cayton2005algorithms}, often referred to as the \emph{manifold hypothesis}. This assumption has rarely been stated rigorously. However, roughly speaking, it posits that high-dimensional real-world data can be sufficiently well-represented by data lying on a lower-dimensional latent manifold, embedded within the feature space (see \citealt{you_ma_manifold, Athanasopouloumanifold, Narayanan_Sanjoy_manifold, Bordt2022manifold, gorban2018manifold, brown2022manifold, Ivanov2021manifold, meiladraft-manifold, Bengiomanifold, izenman2012manifold, Brahma2016}).\footnote{Each individual manifold learning technique also makes a number of further assumptions. However this fundamental assumption, the manifold hypothesis is, by definition, shared by all manifold learning techniques.} 

It will be helpful to distinguish a \emph{local} manifold hypothesis from a \emph{global} manifold hypothesis. Given some dataset, the \emph{local} manifold hypothesis states this dataset can be well-represented by data lying on a lower-dimensional latent manifold, embedded within the feature space. On the other hand, the \emph{global} manifold hypothesis is the proposition that many real-world datasets can be effectively compressed by this kind of manifold learning; indeed that this is a prevalent feature of real-world datasets. One plausible and suitably general way to explicate the local manifold hypothesis could be as follows.\footnote{The main principles of this definition come from \citet{fefferman2016manifold}. They define an algorithm to test the manifold hypothesis within a certain domain, for independent and identically distributed probabilistic data supported on a separable Hilbert Space. For our general purposes, it serves to loosen some of these requirements, whilst restricting ourselves to finite data on an $N$-dimensional space of real-numbers.}

Let $\mathcal{X} \subset \mathbb{R}^N$ be a high-dimensional feature space, with datapoints, $x_i \in \mathcal{X}$. Let $\mathcal{G}_\mathcal{X}(K, V, \tau)$ be the class of sub-manifolds in $\mathcal{X}$ with dimension, $K$, $K$-dimensional volume $\leq V$ and reach $\geq \tau$.\footnote{We only want to consider manifolds above a certain reach and below a certain volume to avoid overfitting; after all, manifolds of sufficiently large volume or low reach could more easily capture every datapoint. Following \citet{fefferman2016manifold}, the reach, $\tau$, of a manifold is defined as the largest distance such that any point within the distance $\tau$ from the manifold has a unique closest point on the manifold. Sometimes loosely described as a measure of smoothness, one can more accurately think of it as a measure of local feature size, related to both local curvature and global \emph{bottlenecks} (see \citealt{manifold-reach} for a more complete explanation). The $K$-dimensional volume is given by the standard Lebesgue measure in $\mathbb{R}^K$. Recall (section \ref{sec:manifold-learning}) that once we define a cost function for some manifold-learning algorithm, we might expect that such an overfitted manifold might nonetheless have a high cost, indicating that it does not properly capture the salient features of the data. In that sense, there is a risk of double-counting this requirement in this definition. One alternative would be to define a cost function from the outset, and require a manifold below a certain cost in the hypothesis. However, here I conservatively choose to stick to the approach used by \citet{fefferman2016manifold}, defining the manifold hypothesis \emph{prior} to specifying any cost function.} Then the manifold hypothesis is the assumption that, for some choice of $K < N, V, \tau$, there exists a manifold, $\mathcal{M} \in \mathcal{G}$ , such that,

\begin{equation}
    \mathcal{L}(\mathcal{M}, \{x_i\}) < \epsilon,
\end{equation}

\noindent where  $\mathcal{L}(\mathcal{M}, \{x_i\})$ is some measure of the average shortest distance (perhaps the mean-squared shortest distance) between the datapoints $\{x_i\}$ and the manifold $\mathcal{M}$, according to some choice of distance (possibly, but not necessarily, the Euclidean distance in $\mathbb{R}^N$), and $\epsilon \in \mathbb{R}$ is some closeness threshold.

We can view the manifold hypothesis as an \emph{data compressibility assumption}. The high-dimensional dataset contains redundancy. As such, the data can be well-represented with the use of a lower-dimensional model, without significant loss of information.

Expressed in this way, the local manifold hypothesis asserts that there exists a manifold $\mathcal{M}$ in $\mathcal{X}$ such that the average distance $\mathcal{L}(\mathcal{M}, P)$ is less than or equal to some specified threshold, chosen based on the desired level of proximity between the data distribution and the manifold. For different applications of the hypothesis, such as with different types of dataset or different manifold learning algorithms, we might choose to consider different manifold parameters, and different ways to measure the average distance and the threshold.

The global manifold hypothesis asserts that this applies to many real-world datasets. A general argument for the hypothesis has not been put forward; however, it has often been presented as a reason why machine learning is possible at all (see \citealt{cayton2005algorithms, colah2014nnmanifold, fefferman2016manifold}). After all, the higher the dimension of the data is, the harder machine learning tasks generally become. The global manifold hypothesis suggests that machine learning algorithms can potentially reduce the complexity of these tasks, by latching onto a smaller number of salient regularities in the data.\footnote{To take a more specific example, this principle is key to explaining the possibility of certain regularization techniques in deep learning, like dropout or weight decay, are effective (see \citealt{srivastava14a_2014, zou_hastie_2005} for further details).} For instance, the task of interpreting handwritten digits in our MNIST dataset is far easier than one might naively fear from the high dimensionality of the data; handwritten versions of the same digits can be summarized by certain common, higher-level features (see \citealt{LLE-MNIST_2017} for one example).

\section{Sloppy Models in the Computational Empirical Sciences}
\label{sec:sloppy}

Manifold learning has usually been applied to machine learning tasks, where we wish to build a simple model of the data for tasks like image recognition. Now let us turn to a framework in the computational empirical sciences,\footnote{I will use the term \emph{computational empirical sciences} to refer to the wide array of scientific disciplines focused on using computational methods to build empirically-supported models of highly complex, high-dimensional target systems. This includes a wide array of fields, including, but not limited to, much of systems biology, chemistry, condensed matter physics, and many areas of engineering and the social sciences.} known as sloppy modeling. Here, the task is to create a simple model of some target system, for the purpose of generating accurate predictions, and hopefully to better understand the system. However, we will see a strong analogy between this framework and machine learning.

In the computational empirical sciences, a model is a function from a real-valued vector of $M'$ parameters to a real-valued vector of $N'$ predictions $f': \mathbb{R}^{M'} \rightarrow \mathbb{R}^{N'}$. Generally, we call $\mathbb{R}^{M'}$ the \textbf{parameter space}, and $\mathbb{R}^{N'}$ the \textbf{prediction space}. Often, the dimensions of the parameter space might represent properties of the system theoretically posited by our model, whereas the dimensions of the prediction space represent observable quantities that we measure. As such, observed datapoints also lie in the prediction space.

We can use measurements to estimate the model parameters. Measurements are represented by a set of real numbered vectors in the prediction space. We write a cost function to measure the distance between the model predictions and the empirical measurements, and tune the model parameters to minimize this cost. If our model is predictively accurate for a given choice of parameters, then the model's predictions should lie close to the measurements. Then we might then use such a model to generate further accurate predictions.

Consider this simple example.\footnote{See \citealt[pages 79-87]{pain2005physics} for a fuller treatment.} Suppose we want to create a computational-scientific model of a pair of apparently identical pendulums, joined together with some string, and starting at rest but with one pendulum displaced (this is our target system). We can measure two things: the displacement of the first pendulum ($x_1$) or the displacement of the second pendulum ($x_2$), each indexed by different times, $t$ (we can think of as $t$ an independent regressor variable). So if we take measurements at ten different times, our measurement space will be $2 \times 10 = 20$ dimensional. Physicists often model such a system as a pair of weakly coupled, identical harmonic oscillators,

\begin{align}
x_1(t) = d \cos\left(\frac{\omega_2 - \omega_1}{2} t\right) \cos\left(\frac{\omega_1 + \omega_2}{2} t\right), \\
x_2(t) = d \sin\left(\frac{\omega_2 - \omega_1}{2} t\right) \sin\left(\frac{\omega_1 + \omega_2}{2} t\right).
\end{align}

\noindent where we have three parameters: $\omega_1$ and $\omega_2$ are normal frequencies of the system, representing the frequency of the pendulums oscillating with the same amplitude  in phase and out of phase, and $d$ is the initial displacement of one pendulum. Note that these natural frequency parameters are theoretical posits of our model, which we do not directly measure. Rather, we infer their values using our model and observations of the displacements.

A model is described as \textbf{sloppy} if its predictions are highly insensitive to most parameter combinations (which we call the sloppy parameter combinations), but are highly sensitive to a small number of parameter combinations (which we call the stiff parameter combinations) (see \citealt{Sloppy_dyson, sloppy_simplicity} for scientific overviews and see \citealt{Freeborn-sloppy} for a philosophical analysis). This allows for significant alterations in the values of sloppy parameter combinations, potentially by factors in the thousands or tens of thousands, with minimal impact on the model's predictive output. Thus a model with $M'$ parameters might operate with a considerably lower \emph{effective} dimensionality in practice. Following \citet{Freeborn-sloppy}, we call a physical target system sloppy if it can be well-represented by a sloppy model, i.e. if we can produce an effective model that is a good description of the system. We could operationalize as the requirement that the datapoints lie close to an effective model manifold.

We can measure the sensitivity of the specific parameter combinations to the observed data using the Fisher Information Matrix (FIM). This gives the expected curvature of the log-likelihood function of the observed data in relation to the model parameters. The eigenvectors of the FIM are termed local or "renormalized" eigenparameters.\footnote{The Fisher Information Matrix gives the expectation of the second-order partial derivatives of the log-likelihood function of the observed data with respect to the model parameters. Viewing this Fisher Information ``matrix'' as a metric (a type (0,2) tensor), it is given by,

\begin{equation}
g_{\mu\nu}(y') = \mathbb{E}\left[ \frac{\partial \log p(x'|y')}{\partial y'^\mu} \frac{\partial \log p(x'|y')}{\partial y'^\nu} \right]
\end{equation}  

\noindent where $\mu, \nu = 1, 2, \ldots, M'$, $y'$ is the $M'$-dimensional parameter vector, $x$ is the $N$-dimensional predictions vector, $p(x'|y')$ gives the likelihood of observing the predictions $x'$ given the parameters $y'$ in the model, and $\mathbb{E}[\cdot]$ gives the expectation with respect to the distribution of the observed measurements.}

 Many real-world systems seem to depend on huge numbers of parameters. However, it becomes increasingly hard to build good models with large numbers of parameters. Just as in machine learning, high-dimensionality can be a major problem in computational scientific modeling. The utility of a sloppy model lies in its ability to effectively capture the salient features of a dataset while demonstrating a robust tolerance to variations in many of its parameters. As such, sloppy systems are suitable targets for \textbf{effective models}, in which some or all sloppy parameter combinations can be ignored. Fortunately, scientists have found that systems across a very wide variety of domains are sloppy, ranging from systems biology to quantum mechanics to particle accelerators \citep{Sloppy_ubiquitous}. Proponents of the sloppy models framework argue that the ubiquity of sloppy systems can help to explain the success of science \citep{Sloppy_dyson, sloppy_simplicity, Freeborn-sloppy}.

Hopefully, by expressing things in this way, the analogy with machine learning is already clear. Note that the scientific model's prediction space, $\mathbb{R}^{N'}$, corresponds to the machine learning model's feature space, $\mathbb{R}^N$. In each case, the dimensions of the space correspond to the real-world observable quantities in the target system; a measurement of each of these quantities corresponds to a datapoints in that space. The scientific model's parameter space, $\mathbb{R}^{M'}$, corresponds to the machine learning model's latent space, $\mathbb{R}^M$. The dimensions of this space correspond to higher-level theoretical quantities (parameters or latent variables) posited by our model of the target system. However, observe that the function $f'$ in the computational empirical sciences takes an opposite direction to the function $f$ in machine learning. The former takes us from our model parameters to the observable predictions, whereas the latter takes us from observable data in the feature space to tune the predicted model parameters. It will often be helpful to assume that such functions are invertible in both the machine learning and computational sciences contexts.

The effective model $m'$ is a function from a simplified, lower-dimensional space of $K' < M'$ effective (or ``renormalized'') parameters, $\mathbb{R}^{K'}$ to the prediction space, $\mathbb{R}^{N'}$. As we will see in there next section, we can also propose a \emph{manifold boundary approximation function}, $g'$, to take us from the high-dimensional, to the low-dimensional prediction space. If the effective model and original sloppy model make the same predictions, then the relation between these functions is shown in figure \ref{fig:category2}. Once again, in reality this is an unrealistic assumption: the two models should give \emph{almost} the same outputs, but some information will be lost in the effective model.

\begin{figure}[htp]
\centering
\begin{tikzpicture}
  \node (RN) at (0,0) {$\mathbb{R}^{N'}$};
  \node (RK) at (6,0) {$\mathbb{R}^{K'}$};
  \node (RM) at (12,0) {$\mathbb{R}^{M'}$};

  \node at (0,1) {\textit{Prediction space}};
  \node at (0,0.5) {\textit{(predictions)}};
  \node at (6,1) {\textit{Effective parameter space}};
  \node at (6,0.5) {\textit{(effective parameters)}};
  \node at (12,1) {\textit{Original parameter space}};
  \node at (12,0.5) {\textit{(model parameters)}};

  \draw[<-] (RN) -- (RK) node[midway,above] {$m'$} node[midway,below] {\textit{Effective model}};
  \draw[<-] (RK) -- (RM) node[midway,above] {$g'$} node[midway,below] {\textit{MBAM}};
  \draw[<-,bend right] (RN) to node[midway,below] {\textit{Original model}} node[midway,below=0.5cm] {$f' = m' \circ g'$} (RM);
\end{tikzpicture}
\caption{A category theoretic representation of the original and effective modeling approaches, assuming that they give the same outputs. Here, $\mathbb{R}^{N'}$ is the prediction space, $\mathbb{R}^{K'}$ is the effective parameter space, and $\mathbb{R}^{M'}$ is the original sloppy parameter space. The functions $f'$, $g'$, and $m'$ represent the original sloppy model, the manifold boundary approximation method (MBAM) function, and the effective model function, respectively.}
\label{fig:category2}
\end{figure}

\section{Manifold Boundary Approximation in the Computational Sciences}
\label{sec:MBAM}

We can derive these effective models by using an information-geometric approach (see \citealt{Sloppy_info_geometry, Sloppy_info_geometry}), in which we endow the model with a little more structure. Each vector of predictions defines a point in the prediction space, $\mathbb{R}^{N'}$, and each vector of model parameters $y'$ generates one such point under the function $f'$. As such (assuming $M' > N'$), we can reinterpret the model as an $M'$-dimensional sub-manifold $\mathcal{R'}$, embedded in the prediction space, $\mathbb{R}^{N'}$. This embedded sub-manifold is defined by the points,

\begin{equation}
 \mathcal{R'}= \{ f(y') \in \mathbb{R}^{N'} : \text{ for all the parameter combinations } y' \in \mathbb{R}^{M'} \}.
 \end{equation}
 
\noindent Here $y'$ gives the manifold coordinates: as such, varying the parameters $y'$ of the model moves along the manifold surface, leading to a different point (vector of predictions) in the feature space in which it is embedded. The collection of all these points (for all possible parameter values) forms the model manifold surface.\footnote{To interpret the model as an embedding, we must make some further assumptions about $f'$. It must be smooth, injective, an immersion, and its underlying continuous function must be a homeomorphism onto its image (see \citealt[pages 21-29]{hirsch1994} for further details).}

The FIM can serve as a Riemannian metric on the model manifold, measuring parameter space distances (in units of standard deviations of the parameter, given their probability distributions under the model). Such distances operationalize the distinguishability between model predictions from different parameter choices. 

We can explore how the model predictions change as we vary corresponding parameter combinations by tracing geodesics along the model manifold. If we move far enough along a geodesic, we may eventually reach a point where further movement would take us to boundaries. Beyond these boundaries, the model's predictions become non-physical, undefined, or irrelevant.\footnote{A $K$-dimensional manifold with boundary $\mathcal{M}$ is a topological space where every point $p$ in $\mathcal{M}$ has a neighborhood homeomorphic to an open set in the Euclidean half-space $\mathbb{R}_{+}^{K} = \{(x_1, \ldots, x_K) \in \mathbb{R}^K : x_K \geq 0\}$. Points in $\mathcal{M}$ that have neighborhoods homeomorphic to an open set in $\mathbb{R}^n$ (the entire Euclidean space) are called \textit{interior points}. Points in $\mathcal{M}$ that have neighborhoods homeomorphic to an open set in $\mathbb{R}_{+}^{K}$ but not in $\mathbb{R}^K$ are called \textit{boundary points}.} For instance, such boundaries can arise when certain parameter combinations are  not physically meaningful, or lead to singularities or mathematically undefined behavior.

Tthe existence of these boundaries on the model manifold represents a general principle of model reduction. This concept suggests that simpler models often arise at the extremes of parameter values, an idea implicit in many areas of physics and elsewhere in the computational sciences. The modern framing of this as 'manifold boundaries' provides a rigorous mathematical foundation for this intuition.

Therefore, geodesic lengths give an indicator of sloppiness: long geodesics correspond to stiff parameter combinations, whilst short geodesics correspond to sloppy parameter combinations, in which the values can be varied over many orders of magnitude without significantly altering the model predictions in $\mathbb{R}^{N'}$. The shape of a sloppy manifold is described as a \emph{hyperribbon}, with many short dimensions and only a few longer dimensions.

There are various ways to identify and utilize these boundaries for model reduction. One method designed for this purpose is the \textbf{manifold boundary approximation method} (MBAM). This uses an information-geometric approach to systematically find lower-dimensional effective models for sloppy systems \citep{Sloppy_reduction}. This allows us to propose a manifold boundary approximation function, $g' : \mathbb{R}^{M'} \rightarrow \mathbb{R}^{K'}$, taking vectors of the original parameters to corresponding vectors of the effective parameters. The MBAM algorithm proceeds as follows to find a particular boundary of the model manifold.

\begin{itemize}
    \item We begin with the dataset with datapoints ${x'_i} \in \mathbb{R}^{N'}$, and the embedded model manifold, $\mathcal{R'}$. We posit that the system is sloppy.
    \item The goal is to find an embedding function $g' : \mathbb{R}^{M'} \rightarrow \mathbb{R}^{K'}$ that projects each high-dimensional parameter vector in the parameter space onto the $K'$-dimensional manifold, $\mathcal{M}$. We seek an effective model, $m':\mathbb{R}^{K'} \rightarrow \mathbb{R}^{N'}$, such that for each parameter vector within some chosen domain, $y'_i$, $f'(y'_i) \approx m'(g'(y'_i))$, i.e., the sloppy model and the effective model should make approximately the same predictions within this domain.
    \item We find an embedding, $g'$, as follows:
        \begin{itemize}
            \item \textbf{Model Fitting:} Initially, fit the model to data to find a best-fit point in the parameter space using an appropriate cost function,
                \begin{equation}
                    C:\mathbb{R}^{N'} \times \mathbb{R}^{M'} \rightarrow \mathbb{R},
                \end{equation}
                \noindent which assigns a real number to each pair of points, one from the prediction space and one from the parameter space.\footnote{The task of fitting the model parameters can be interpreted as projecting the data onto the model manifold \citep{sloppy_simplicity}.}
            \item \textbf{Eigenvalue Analysis:} Compute the FIM at this point and perform an eigenvalue analysis. Identify the direction associated with the smallest eigenvalue, which corresponds to the sloppiest parameter combination.
            \item \textbf{Geodesic Tracing:} Trace a geodesic in the parameter space along this sloppiest direction. This path leads to a boundary of the model manifold where the insensitive parameter becomes redundant.
            \item \textbf{Model Reduction:} At the manifold boundary, effectively remove or fix the redundant parameter, thereby reducing the model's dimensionality.
            \item \textbf{Effective Model Fitting:} The effective model parameters are fit to the data using an analogous cost function, 
                \begin{equation}
                    C:\mathbb{R}^{N'}\times \mathbb{R}^{K'} \rightarrow \mathbb{R},
                \end{equation}
                \noindent which assigns a real number to each pair of points, one from the prediction space and one from the effective parameter space, designed to measure how well a map preserves salient geometric and topological features of the data. 
            \item \textbf{Iteration:} Repeat the process as needed to simplify the model further, focusing each time on the next sloppiest direction.
        \end{itemize}
    \item Finally, the effective parameter vectors, $y'_i$, are represented in the lower-dimensional effective parameter space by their images under the embedding, $g'(y'_i) = y'_i \in \mathbb{R}^{K'}$.
\end{itemize}

Recall the problem with overfitting in manifold learning discussed in section \ref{sec:manifold-learning}. The MBAM procedure is generally resistant to such overfitting. Unlike manifold learning techniques, which might find arbitrary lower-dimensional representations to fit data, MBAM is constrained to move along existing model structures. The boundaries it finds correspond to limiting cases of the original model, preserving its core structure rather than introducing new, potentially overfitting parameters. This process is guided by the model's intrinsic geometry, not by fitting to specific data points. Thus MBAM does not introduce new complexity to match particular observations.

\begin{figure}[htbp]
    \centering
    \begin{subfigure}{0.95\textwidth}
        \centering
        \includegraphics[width=\linewidth]{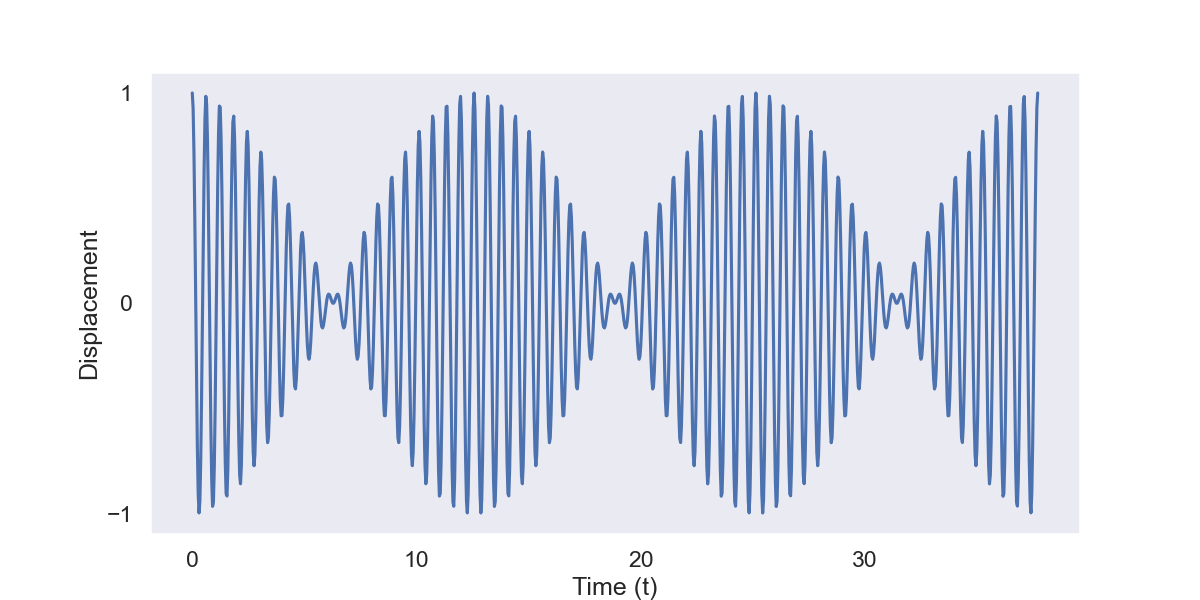}
        \caption{Predicted displacement of the first pendulum according to the original model, for some choice of parameter values.}
        \label{fig:beatsa}
    \end{subfigure}%

    \begin{subfigure}{.95\textwidth}
        \centering
        \includegraphics[width=\linewidth]{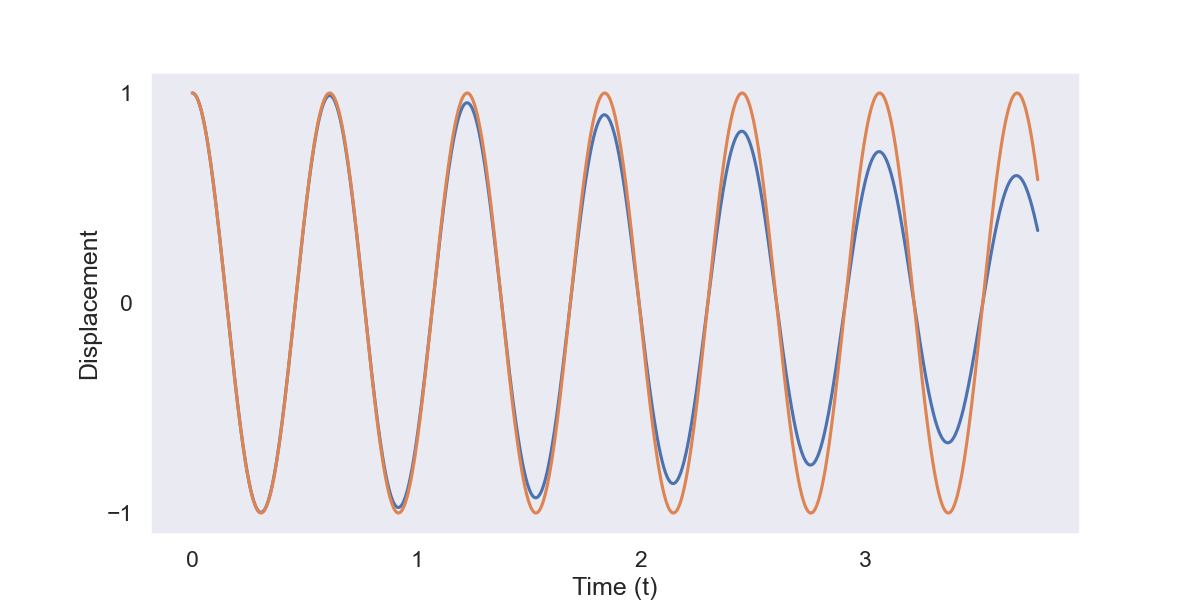}
        \caption{Predicted displacement of the first pendulum with the same parameter values,  at small times, according to the original model (\textcolor{blue}{blue}) and effective model (\textcolor{darkorange}{orange}). Notice that the two models almost agree at sufficiently small times.}
        \label{fig:beatsb}
    \end{subfigure}
\end{figure}

To illustrate a simplified, toy version of this procedure, consider our coupled pendulums model once more. This model is not sloppy \emph{in general}; however, it can illustrate some principles effective model building because some of its parameter combinations can become sloppy in certain domains, such as when the times $t$ are very small. Here, we shall focus on just the displacement of the first pendulum, $x_1$. First, let us rewrite the model in terms of some new parameter combinations, $\omega_\text{h} = \omega_2 + \omega_1$ and $\omega_\text{l} = \omega_2 - \omega_1$. Now the displacements $x_1$ and $x_2$ are characterised by a higher frequency ($\omega_\text{h}$) modulated sinusoidal oscillation, varying within a lower frequency ($\omega_\text{l}$) sinusoidal envelope (figure \ref{fig:beatsa}). At small enough times, $t$, we might find that the model predictions are highly insensitive to the value of $\omega_l$; only the high frequency variations seem to matter: in this domain, $\omega_l$ has become sloppy. We might find that the geodesics reveal a model boundary at $\omega_l \rightarrow 0$. At this boundary, corresponding to small times and small couplings, we can eliminate the dependence on $\omega_l$, and rewrite our model, 

\begin{align}
x_1(t) = d  \cos\left({ \omega_\text{h}} t\right),
\end{align}

\noindent which we can think of this as a effective model of the system adequate at a particular domain, with just two parameters, $d$ and $\omega_\text{h}$ (figure \ref{fig:beatsb}).

\section{Manifold Boundary Approximation and Manifold Learning}

Prima facie, MBAM appears to be a fundamentally different procedure to Manifold Learning, as described in section \ref{sec:manifold-learning}. There is some truth to this. After all, manifold learning provides a way to build a simplified model of the data, taking us from datapoints, ${x_i}$, in the data space, $\mathbb{R}^N$, to prediction vectors, ${y_i} = m(x_i)$,  in a reduced latent space, $\mathbb{R}^K$. On the other hand, the Manifold Boundary Approximation Method takes us from parameter vectors ${y'_i}$ in the parameter space, $\mathbb{R}^{M'}$ to effective parameter vectors, ${y'_i} = g'(y'_i)$, in the effective parameter space, $\mathbb{R}^{K'}$. Recalling our analogy between machine learning and the computational empirical sciences, the data space corresponds to the prediction space, the latent space corresponds to the parameter space, and the reduced latent space corresponds to the effective parameter space. As the diagrams in figures \ref{fig:category} and \ref{fig:category2} demonstrate, these are not the same procedure.\footnote{Indeed, they have generally been viewed as two essentially distinct kinds of procedure \citep{sloppy_simplicity, sloppy_recent, sloppy_kullback}.}

However, recall that before starting the MBAM procedure, we reinterpreted the model $f'(y'_i) = x'_i$ as an $M'$-dimensional manifold, $\mathcal{R'}$, with $y'$ giving the manifold coordinates, embedded in the prediction space, $\mathbb{R}^{M'}$. We begin the MBAM procedure by finding the parameter vector on this manifold, best tuned to the datapoints, $x'_i$, and then follow a geodesic to reach the manifold boundary. Finally, the sloppy parameter combination is eliminated, and the effective model is re-tuned to the data. In effect, this procedure identifies a $K'$-dimensional sub-manifold of the dataspace, $\mathbb{R}^{N'}$, that is tuned to the data, just as with the case of manifold learning.

To be clear, the procedures are not identical here. Under manifold learning, we find the specific sub-manifold (for a given $K$) that is best tuned to the data according to the cost function. By contrast, the manifold boundary found by following a sloppy geodesic is in no sense guaranteed to be the $K'$-dimensional sub-manifold best tuned to the data according to any given cost function. Rather than simply opting to find the best sub-manifold, in the MBAM procedure, we first seek to eliminate the sloppiest parameter combinations. These are precisely the parameter combinations that are least sensitive to the data, and in general likely to be hardest, and least relevant, to tune. The purpose of eliminating such parameter combinations is precisely to make the task of tuning the effective model easier. Whilst MBAM does not entail finding the sub-manifold most tuned to the data, it does find the sub-manifold whose most \emph{stiff} parameter combinations are best tuned to the data.

Here it is worth considering the two procedures epistemically. When performing manifold learning, we begin with knowledge of the data, $x_i \in \mathbb{R}^N$. We do not assume knowledge of the prior model $f$ or of the latent variables we wish to measure that make up $\mathbb{R}^M$. The manifold learning procedure uses the data to give us the function $m$ that can best represent the datapoints (and in effect the function $g$), and the reduced latent space, $\mathbb{R}^N$.  Now consider MBAM. We begin with analogous knowledge, of the data, $x'_i \in \mathbb{R}^{N'}$, but also require prior knowledge of the model function, $f'$ and the model parameters that make up $\mathbb{R}^{M'}$, . The manifold learning procedure uses these to give us the function $g'$ (and in effect the function $m'$ that allows us to best represent the datapoints using the stiff parameters) and the effective parameter space, $\mathbb{R}^{K'}$. 

Thus, both procedures construct the same kind of object, a sub-manifold of the feature space or prediction space, $\mathbb{R}^N$ or $\mathbb{R}^{N'}$, tuned appropriately to the datapoints $x_i \in \mathbb{R}^N$ or $x'_i \in \mathbb{R}^{N'}$. However, there are two key epistemic differences. First, when performing MBAM, we start with greater knowledge: we must already have some prior model of the data, $f'$, unlike in manifold learning. The epistemic role of $f'$ is precisely to help us identify the stiff parameter combinations in constructing $\mathbb{R}^{K'}$. Further note that identifying these is the \emph{only} epistemic role of the model $f'$: the effective parameter tuning takes place with respect to the data, in precise analogy to manifold learning. Second, when we perform MBAM, we restrict ourselves to best representing the data only using these stiff parameter combinations.

To stress this point, let us consider the MNIST dataset example once more. Suppose that our machine learning specialist has finally found a model, $f$, taking the real-valued vector of 784 pixles in each time to a vector of ten-real-valued output classifications. However, they find the model overly complex and difficult to tune, due the many apparently insensitive parameters. Perhaps the task would be easier if they could find a reduced dimensionality model, which doesn't depend on all these parameters. They use the data to identify the insensitive parameter combinations, and then build a effective model of the data, $g: \mathbb{R}^K \rightarrow \mathbb{R}^N$. Observe how this process relates closely to the process of building a dimensional reduction model, described in section \ref{sec:manifold-learning}, albeit utilising already-existing model.

Confusion is natural given the different ways functions have been defined in machine learning and the computational empirical sciences. However, whilst the methods differ, it perhaps makes sense to think of the MBAM procedure as akin to a special kind of manifold learning procedure, in which a prior model, $f'$ is used to identify the stiff parameter values, and then the reduced latent manifold (or effective parameter space) is fit to the data only using the stiff parameter combinations.

This motivates a reconsideration of \emph{sloppiness} itself. Following \citet{Sloppy_info_geometry2, Sloppy_reduction}, we can describe a model $f':\mathbb{R}^{M'} \rightarrow \mathbb{R}^{N'}$ as sloppy if $\mathcal{R'}$ (the corresponding $M'$-dimensional sub-manifold of $\mathbb{R'}$) has few stiff geodesics, and many sloppy geodesics. Each such sloppy geodesic must have at least one manifold boundary. The MBAM algorithm provides one way to identify and utilize these boundaries, allowing us to eliminate sloppy parameter combinations and build an effective model of reduced-dimensionality. If we can build an effective model with $ M'-K'$ fewer parameters, then this model will correspond to a $K'$-dimensional sub-manifold of $\mathbb{R}^{N'}$.


So, one plausible explication of sloppiness would be with the following criterion. Let $\mathcal{X'} \subset \mathbb{R}^{N'}$ be a high-dimensional prediction space, with datapoints, $\{x'_i\} \in \mathcal{X'}$. Require that $\mathcal{Y'} \subset \mathbb{R}^{M'}$ is a high-dimensional parameter space, with an embedding $f' : \mathcal\mathbb{R}^{M'} \rightarrow \mathbb{R}^{N'}$ defining a sub-manifold in the prediction space, $\mathcal{R'} \subset \mathbb{R}^{N'}$. Let $\mathcal{G'}_\mathcal{X'}(K', V, \tau)$ be the class of \emph{effective model} sub-manifolds in $\mathcal{R'}$ with dimension, $K'$, volume $\leq V$ and reach $\geq \tau$. Then the sloppiness criterion demands that, for some choice of $K' < N', V, \tau$, there exists an effective manifold, $\mathcal{M'} \in \mathcal{G'}$ , such that,

\begin{equation}
    \mathcal{L}(\mathcal{M'}, \{x'_i\}) < \epsilon,
\end{equation}

\noindent where  $\mathcal{L}((\mathcal{M}, \{x_i\})$ is some measure of the average shortest distance between the datapoints $\{x'_i\}$ and the manifold $\mathcal{M}$, according to some choice of distance, and $\epsilon \in \mathbb{R}$ is some closeness threshold.

Observe that this sloppiness criterion is essentially the manifold hypothesis from section \ref{sec:manifold-hypothesis}, alongside an addition requirement: the sub-manifold that describes the data must be an effective model manifold, produced from a sloppy model. This additional requirement restricts the scope of the models under consideration, but in a reasonable way: finding an effective model precisely involves eliminating parameter combinations which are least sensitive to the data. 

Another way to think of this is that manifold learning might be a \emph{more} theory-independent way of building a dimensional reduction model of the data. However, as noted, any particular manifold learning algorithm will add additional assumptions. Finding an effective model is \emph{more} theory-dependent: we begin with an assumption that a prior model $f'$ can offer a good description of the system, but has some superfluous (sloppy) parameter combinations which can be eliminated, reducing the dimensionality of the model. However, above this assumption, the additional assumptions of effective-model building are quite weak, namely that the improved model of the system involves precisely removing those irrelevant parameter combinations.

We could imagine MBAM's goal as reducing the number of parameters, while retaining as much relevant information from the original model as possible, ensuring that the essential features and behaviors of the system are preserved. We could imagine a ``cost function'' in terms of trace of the FIM, to measure the fidelity of the reduced model to the original model’s manifold. Against such a cost, then MBAM would behave like a \emph{greedy algorithm}\footnote{A greedy-algorithm iteratively makes the locally optimal choice at each step, typically without backtracking. Such an algorithm may not find a global optimum.}, making locally optimal choices at each step without guaranteeing a globally optimal solution. As such, MBAM does not explore the whole space of solutions.

We could consider other algorithms  for achieving this goal. At the most constrained end of the spectrum, MBAM represents a greedy approach, in which it finds a particular $K'$-dimensional boundary through iterative geodesic tracing. While computationally efficient, this does not guarantee finding the globally optimal reduced model. By contrast, manifold learning seeks to find the best K-dimensional representation of the data without constraints on the form of that representation, and requires explicit smoothness assumptions to avoid overfitting. Less constrained alternatives to MBAM might be algorithms that find the optimal $K'$-dimensional edge or boundary of the model manifold, or even choosing a $K'$-dimensional sub-manifold of the original model manifold, without the requirement that it be an edge or boundary. Such algorithms could still inherit structural constraints on the model, and therefore might not require a further procedure to prevent overfitting. Such algorithms could plausibly have useful properties relevant to both scientific model-reduction and machine learning.

Just like the manifold hypothesis, we can view sloppiness as a \emph{compressibility assumption}. The high-dimensional dataset and the model contain redundancy. As such, the data can be well-represented with the use of a lower-dimensional effective model, without significant loss of information.

\section{Effective Field Theories in Physics}
\label{sec:renormalization}

The sloppy models program is closely tied to the modern effective field theories (EFTs) program in quantum field theory and particle physics.\footnote{Indeed, at least some cases of effective field theory building can be understood as instances of effective model building in sense meant in sections \ref{sec:sloppy} and \ref{sec:MBAM} (See \citealt{Sloppy_RG, Sloppy_RG_info, Sloppy_dyson} for further details and \citealt{Freeborn-sloppy} for a philosophical discussion.} It is therefore worth considering to what extent the effective field theories program in physics can also be understood through the use of manifold learning methods.

EFTs are built on the principle that quantum field theoretic phenomena decouple at different energy or distance scales. This means that to describe physics at a certain scale, it is only necessary to consider relevant degrees of freedom and interactions at that scale. Furthermore, given our ignorance about physics at certain scales, it is often necessary to do so. As with sloppy models, EFTs simplify the description of complex systems by identifying the most relevant degrees of freedom and parameters at a given scale.\footnote{For further details see \citealt*{Binney_RG}, \citealt{Weinberg} and \citealt*{Duncan}. For a philosophical overview, see \citealt*{Butterfield_Intro} and \citealt{Butterfield_Technical} for further details).}

Many highly successful theories are effective field theories, including Fermi's Theory of Weak Interaction, Chiral Perturbation Theory, Heavy Quark Effective Theory, and Ginzburg-Landau Theory. Indeed, on the modern approach pioneered by \citet{Wilson}, we should expect \emph{every} quantum field theory to be an effective field theory, including the Standard Model of particle physics.

We can think of a quantum field theory as a scientific model along essentially the same lines as the computational models in section \ref{sec:sloppy}. We can define a space of predictions, $\mathbb{R}^{N'}$, consisting of measurable quantities, perhaps including physical quantities such as scattering amplitudes, cross-sections or decay rates; and a space of parameters, $\mathbb{R}^{M'}$ perhaps including the field variables and their so-called \emph{bare} coupling constants, such as masses and charges.\footnote{We might think of both the field variables \emph{and} field parameters as parameters of the theory. The theory posits both the \emph{kinds} of fields (for example, scalar, vector, spinor etc), expressed through the field variables, and their bare coupling constants.} We can interpret the theory as a function, $f':\mathbb{R}^{M'} \rightarrow \mathbb{R}^{N'}$, mapping from a selection of parameters to a set of predictions. The theory is usually written in terms of a Lagrangian, $\mathcal{L}(\phi_i, \lambda_j)$, where $\phi_i$ are posited field variables, and $\lambda_j$ are the field parameters. We can derive the equations of motion, and eventually the predictions about observables, through a rather involved process, starting with the Lagrangian.

Unfortunately, calculations using non-trivial interacting quantum field theories are typically found to lead to problematic mathematical divergences. To tackle these divergences, physicists modify these theories through a family of correction techniques known as \textbf{regularization}. For example, a simple way to do this is to impose a momentum cutoff scale, at a much higher energy than the interactions we wish to study. These corrections can render the theory finite, but usually lack a principled physical motivation. 

The solution, \textbf{renormalization}, involves adjusting the bare parameters of the theory to remove dependence on any regularization scale. The resulting \emph{renormalized theory} has new \emph{renormalized parameters} that shift with the energy or distance scale, at which we describe the theory. We call the shifting of these renormalized \emph{running} coupling constants the \textbf{renormalization group flow} (RG flow) through the parameter space. We derive differential equations to describe how the renormalized couplings must vary with the scale, if we impose the requirement that the physical observables in the prediction space must remain the same. 

Crucially, the parameter trajectories under RG flow will remain \emph{within} the parameter space of the theory. These parameter trajectories often lead to fixed points or surfaces,\footnote{There are generally two kinds of fixed points – Gaussian (or free-field) fixed points, where interactions vanish, and non-Gaussian fixed points, where the interactions reach a non-zero constant value. These fixed points are also crucial in understanding critical phenomena in phase transitions.} at which the value of the renormalized coupling constants cease to change with the scale. Many different theories, those with the same fields and symmetries may flow towards the same fixed surface.\footnote{For a proof in the case of one simple scalar theory, see \citealp{Polchinski1984}.} As such, the fixed surfaces are said to define \emph{universality classes} of theories that share the same behavior at some scale.

Hence, near these fixed regions, the theory can be thought of as exhibiting a kind of self-similar behavior across scales, an approximate scale-invariance. In effect, the renormalizable part of the theory can be approximately decoupled from the physics energy scales. A linear approximation of the RG flow equations near stable fixed points reveals a small number of unstable directions, in contrast to the majority that are stable. Unstable directions correspond to parameters for which small changes can result in large changes to the theory's predictions. These correspond to \textbf{relevant} and \textbf{marginal} parameters. Stable directions correspond to parameters for which small changes only lead to small changes to the theory's predictions. These correspond to \textbf{irrelevant} parameters.\footnote{Roughly speaking, relevant parameters are those that whose effect on the theory's predictions increases close to a stable fixed point. Irrelevant parameters are those that whose effect on the theory's predictions decreases close to a stable fixed point. Other parameters are described as marginal (see  \citet[pages 245-246]{Goldenfeld1992} for further details).} Therefore, the predictions of the theory become dominated by a smaller number of \emph{relevant} and \emph{marginal} parameters.

In consequence, we can construct a lower-dimensional \textbf{effective field theory} by eliminating the irrelevant renormalized parameters. Such effective field theories provide a good model of the system at certain energy scales, usually at low energy, but break down at other scales, often high energy scales. As such, we can construct predictively successful, effective low-dimensional, low-energy theories even whilst remaining ignorant of the physics at higher energy scales.

Let us consider a simple example, an imaginary, simple scalar quantum field theory with two interacting fields, $\phi_L$ and $\phi_H$ with different masses, $m_L$ and $m_H$ respectively, with $m_L \ll m_H$. Suppose that we want to find a low-energy effective theory, relevant to sclaes $\Lambda \ll m_H$. We summarize the original theory with the Lagrangian,

\begin{equation}
\mathcal{L} = \frac{1}{2}(\partial_\mu \phi_L)^2 - \frac{1}{2}m_L^2\phi_L^2 +
              \frac{1}{2}(\partial_\mu \phi_H)^2 - \frac{1}{2}m_H^2\phi_L^2 
              - \lambda \phi_L^2 \phi_H^2,
\end{equation}

\noindent where $\partial_\mu$ represents the partial derivative with respect to spacetime coordinates and $\lambda$ is the coupling constant for the interaction between the two fields, $\phi_L$ and $\phi_H$. At low energies, we find that the parameters associated with high-mass particles become irrelevant, and the lower mass fields become effectively decoupled from them. The Lagrangian relates to the field configurations of the theory by means of the partition function,

\begin{equation}
    Z = \int \mathcal{D}\phi_L \mathcal{D}\phi_H \, e^{i\int d^4x \, \mathcal{L}(\phi_L, \phi_H)}.
\end{equation}
    
\noindent We can eliminate the higher mass degrees of freedom by integrating over them and defining a new effective Lagrangian, $\mathcal{L}_\text{eff}$,  as follows.\footnote{There are two essential steps to this process. First, the Applequist-Carrazone decoupling theorem shows that the degrees of freedom associated with high-mass particles are suppressed at low energies (see \citealt{Appelquist_Carrazone_decoupling} for further details) Second, the degrees of freedom can be separated and integrated out, to form an effective field theory (see \citet{Wilson, Wilson_decoupling} for further details).} 

\begin{equation}
    Z = \int \mathcal{D}\phi_L  \, e^{i\int d^4x \, \mathcal{L}_\text{eff}(\phi_L)}.
\end{equation}

\noindent Unfortunately, in general such an effective Lagrangian may be characterized by an infinite series of terms,

\begin{equation}
\mathcal{L}_{\text{eff}} = \frac{1}{2}(\partial_\mu \phi_L)^2 - \frac{1}{2}m_L^2\phi_L^2 + \sum_{n=1}^{\infty} \frac{c_n}{m_2^n} \mathcal{O}_n(\phi_1),
\end{equation}

\noindent where $c_n$ are coefficients that depend on the details of the full theory, including the coupling constant, $\lambda$. Fortunately, in this case, power-counting considerations\footnote{The operators built from fields (in this case, $\phi_L$) can be organized according to the size of their contribution in a systematic expansion. See \citet[pages 51-81]{Burgess2020} for further details.} and renormalization group arguments can show that these terms become irrelevant in the low energy limit we are interested in, resulting in a well-defined low-energy effective field theory, with Lagrangian $\mathcal{L}_{\text{eff-low-energy}}$,

\begin{equation}
\mathcal{L}_{\text{eff}} = \frac{1}{2}(\partial_\mu \phi_L)^2 - \frac{1}{2}m_L^2\phi_L^2.
\end{equation}

\noindent In this case, the low-energy theory looks exactly as we might have expected, with only the free fields for the low mass field, and no coupling to the high mass field. At this energy scale ($\Lambda \ll m_H$) and the high mass field is \emph{frozen out}. The direct production of heavy particles associated with the high mass field energetically unfeasible, and their indirect effects, such as contributions to quantum corrections, are also negligible. Our effective model of the system describes free, low-mass fields.

Clearly, we can understand this overall procedure of finding a lower-dimensional effective field theory as a form of dimension reduction of our original theory's parameter space. Observe how similar this approach was to finding an effective model of the coupled oscillators in sections \ref{sec:sloppy} and \ref{sec:MBAM}.

\section{Renormalization and Compressibility}
\label{sec:renormalizability}

For some simple models, renormalization group flow towards fixed regions can be recovered as a special case of MBAM. Indeed, just as the manifold hypothesis has been proposed as a reason why machine learning is possible, and sloppy modeling has been credited as an explanation for the success of science, many cite the renormalization group procedure as an explanation for successful theory building in high energy physics \citep{Wallace_naivete, PorterWilliams, Fraser_Realism, Fraser_RealProblem, Fraser_Towards, Miller, Weinberg}. This raises the question: can we say something stronger about the relationship between effective theory building in renormalized theories and the other manifold learning procedures we have discussed so far?

Here, it will serve to step back and consider the renormalization group in a more general setting than just quantum field theory. Recall from section \ref{sec:renormalization}, that the renormalization group involved transforming parameters within the parameter space under a change of energy or some other scale, without changing the theory's predictions in the prediction space. The renormalization group transformation can be understood as a coarse-graining procedure, in which the short-distance degrees of freedom of the model are integrated out, effectively viewing the system with less and less precision. In order to keep the model predictions in agreement, these scale transformations require us to transform between points in the parameter space \emph{and}, correspondingly, to rescale the prediction space.

Like the MBAM procedure, RG flow does not generally face overfitting problems of the type we discussed in section \ref{sec:manifold-learning}. Unlike data-driven dimensional reduction techniques, which might find arbitrary lower-dimensional representations to fit data, RG flow is constrained by the structure of the underlying theory. The transformations involved in RG flow are guided by the symmetries of the system and the requirement of scale invariance at fixed points. These transformations require that the effective models bear a self-similarity in form to the original model (except for a possible reduction of parameters). Ideally, each step in the RG flow corresponds to a physically meaningful transformation of the theory, such as integrating out short-distance degrees of freedom to obtain an effective theory. Once again, the process is guided by the theory's structure, not by fitting to specific data points.

Consider the following simple example (originating with \citealp{Kadanoff_1966}). In the one-dimensional zero-field Ising model, we consider an infinite chain of coupled spins. A single parameter, $J$ gives the coupling between neighbouring spins; the spins $s_i$ at sites indexed by $i$, take values of $\pm 1$ and give the predictions of the model. The model is often summarized using the Hamiltonian,

\begin{equation}
    H = -J \sum_{i} s_i s_{i+1}
\end{equation}

We can coarse-grain or ``renormalize'' the model by averaging out short-range details and focusing on long-range behavior. One way to do this is through the Kadanoff \emph{block spin transformation}, in which we group spins into blocks of spins and then sum over the spins within each block to define new, effective spin variables. If we use blocks of 2 spins, the averaged spin of the block could be,

\begin{equation}
    S_{\text{new}} = \text{sign}(s_1 + s_2).
\end{equation}

\noindent After this process, we effectively reduce the number of degrees of freedom in our prediction space: we are now sensitive to only half as many spins as before. Perhaps now, we keep only the predictions for the spins at even sites in our prediction space. This might inspire us to write a new effective model of the system, in which the effective Hamiltonian, $H'$, will only involve these spins,

\begin{equation}
H' = -J' \sum_{i} s_{2i} s_{2i+2},
\end{equation}

\noindent and in which the new effective coupling, $J'$, serves as an effective coupling between what were previously blocks of spins. Requiring that the new model spin predictions correspond to the predictions of the original model, and assuming a probability distribution over the spins, we can derive a functional form of $J'$ in terms of J, and derive renormalization group flow equations to understand how the coupling changes with scale. Successive iterations of the transformation lead towards fixed points, in which the effective couplings do not change further. Observe then that the transformation of the parameter space, whilst keeping the effective predictions in correspondence with the previous predictions necessitates a loss of sensitivity to some of the degrees of freedom in the prediction space. In a sense, the effective model does not map to as many predictions.

More generally, assuming the computational modeling framework from sections \ref{sec:sloppy} and \ref{sec:MBAM} it will help to define the renormalization group flow as any transformation on the embedded model manifold, $\mathcal{R'}$, or within the parameter space $\mathbb{R}^{M'}$ under some change of scale, which coarse grains the model's predictions. Renormalization group transformations therefore require that the effective models bear a  self-similarity in form to the original model (except for a possible reduction of parameters). As such, any coordinate-invariant geometric and topological features of the model manifold will remain fixed. This raises a question: if the model manifold does not change, and renormalization simply constitutes a flow along its surface, how is it that the renormalized model will lose degrees of freedom in the prediction space, $\mathbb{R}^{N'}$?

The key is to realize that the manifold's metric, given by the FIM, representing the distinguishability between model predictions from different parameter choices decreases. \citet{Sloppy_RG_info} show that we can quantify the loss of information from discarding these degrees of freedom discarded through coarse-graining by finding how the metric tensor changes under a coarse-graining application. Let us specify a continuous coarse-graining procedure, where as the smallest length scale, $ l = l_0 \exp(b)$, changes, the parameters change according to $ \frac{dy'^\mu}{db} = \beta_\mu$, where $y'^\mu$ are the parameters, and $\beta $ are the beta functions, which define the flow so as to preserve the predictions. Then the change in the metric under this flow is given by a modified Lie derivative, $\mathcal{L}_\beta$,

\begin{equation}
    \mathcal{L}_\beta g_{\mu\nu} = \beta^\alpha \partial_\alpha g_{\mu\nu} + g_{\alpha\mu} \partial_nu \beta^\alpha + g_{\alpha \nu} \partial_mu \beta^\alpha - L \partial_L g_{\mu\nu},
\end{equation}

\noindent where the first term represents the directional change of the metric, and the second and third terms represent the change in the parameter space distances, as the parameters shift. The fourth term arises from the coarse-graining of the model, assuming that the size of the observed system length $L$ shinks, according to $\frac{dL}{db} = -L$.\footnote{As L is not a parameter, we must supplement the usual Lie derivative with this fourth term.}  \citet{Sloppy_RG_info} find that the metric decreases along the irrelevant directions, whilst it is preserved along the relevant and marginal directions.\footnote{This observation has been corroborated by computer simulations \citep{Sloppy_RG}. Note that, in a closely related model, \citep{sloppy_crucial} demonstrate that the changes in the metric due to the model deformation exactly corresponds to the changes in the metric induced by parameter flow.}

This motivates a careful consideration of the renormalization group flow procedure. This procedure is reminiscent to MBAM, but with two key differences. First, during renormalization group flow, we do not necessarily travel along the geodesics corresponding to the sloppiest parameter combinations. More crucially, we have added the additional step of course graining the predictions of the theory. This yielded a decrease in the FIM metric in certain directions, effectively increasing the sloppiness of the theory. Indeed, the parameters of such systems have been found to become increasingly sloppy as they approach fixed points under renormalization group flow  \citep{Sloppy_RG, Sloppy_RG_info}. This increase in sloppiness corresponds to a loss of information as we perform renormalization group flow, precisely corresponding to the loss in sensitivty to certain degrees of freedom in the prediction space.\footnote{For an alternative, but potentially related sense in which the renormalization group flow corresponds to a loss of information, see \citet{Zomolodchikov}.} 

Therefore, one way to understand the renormalization group flow procedure as we have defined it, is as a particular kind of transformation within the parameter space that also coarse grains the prediction space, thereby modifying the model to increasing sloppiness. It is precisely those irrelevant and sloppy parameters that we remove when building an effective theory. Insofar as effective theory building in physics involves increasing model sloppiness and then creating an effective model in the sense of section \ref{sec:MBAM}, then this too could be interpreted as akin to a special kind of manifold learning. If we understand the renormalization group procedure as increasing sloppiness, then the task of effective theory building is straightforwardly analogous to the construction of an effective model of an (at least somewhat) sloppy system. This kind of effective theory construction seems to depend upon a criterion of theoretical compressibility. 



\section{Conclusions}
\label{sec:conclusions}


Manifold learning, the sloppy models program, and effective field theories operate in three different scientific domains. Nonetheless, there are strong analogies between the three fields. Of course, all three share a basic principle in common: they seek to reduce the dimensionality of some data, model, or theory. However, more fundamentally, MBAM shares a close analogy with manifold learning. The two techniques have generally been seen as fundamentally different: manifold learning begins with a high-dimensional dataset and seeks to produce a low-dimensional model of it, whereas MBAM also seeks to build a lower-dimensional effective model of an already-existing high-dimensional model. However, both ultimately produce a low-dimensional model of the data: the difference is that MBAM uses greater prior knowledge to do so, in particular using a prior model to help identify the sloppy parameter combinations to remove. As such, I have argued that MBAM can be viewed as akin to a special kind of manifold learning.

Likewise, effective theory building in physics bears a close relationship to manifold learning. The renormalization group procedure can be understood as being in some ways analogous to MBAM, transforming the parameters of the theory. However, by simultaneously applying a coarse-graining to the predictions of the theory, it more drastically transforms the model, increasing the model's sloppiness. As such, effective theory building in physics could also be understood as akin to a special kind of manifold learning.

The manifold hypothesis underpins large areas of research in machine learning. If the global manifold hypothesis is right, then it may contribute to an explanation of why machines are capable of learning from complex data. Likewise, the sloppiness of real-world systems, or the existence of fixed points under renormalization group flow may contribute to explanations of why we can build scientific models of highly complex real-world systems. I have argued that all these assumptions share a basic common form: though the systems in question are superficially complex, they contain redundancy in the form of regularities. As such, the systems can be \emph{compressed}. We can construct lower-dimensional effective models of the system by latching onto these regularities.

These technical connections between manifold learning, sloppy models, and effective field theories may have implications for several key philosophical debates, especially regarding the renormalization group. Recall that Batterman argues that some renormalization group phenomena are not reducible to lower-level theories. He contends that renormalization group techniques reveal how macroscopic properties emerge from microscopic interactions in ways that resist traditional forms of reduction due to the need for idealizations, such as the thermodynamic limit or infinite size assumptions.  Viewing these methods through the lens of dimensional reduction techniques described in sections \ref{sec:dimensional}, \ref{sec:manifold-learning}, and \ref{sec:MBAM} may help to shed some light. Each technique involves idealizations such as the successive elimination of sloppy parameter combinations at manifold boundaries in the MBAM procedure, or the reduction in dimensionality from the feature space to the latent space in manifold learning.  These idealizations are cases of the kinds of infinite idealization discussed in \citet{Batterman2002, Batterman2005, Batterman2011}: we let certain parameters or parameter combinations shrink indefinitely small, to capture essential behaviors of physical systems.

Nonetheless, such dimensional reduction techniques seem to take the explicit form of approximate reductions, akin to that suggested by \citet{Butterfield_Intro} in the context of the renormalization group.\footnote{Recall that Butterfield argues that, in the context of the renormalization group, reduction and emergence are not mutually exclusive, suggesting that even phenomena that appear emergent can often be reconciled with a form of reduction when idealizations are understood as approximations rather than ontological separations.} In a traditional reduction scheme \citep{Nagel, Schaffner} two key conditions must be met: there must be bridge laws that systematically correlate the theoretical terms of both theories, and the laws of the reduced theory must be logically derivable from the reducing theory's laws combined with the bridge laws and any necessary auxiliary assumptions. The dimensional reduction schemes discussed in this paper seem to take precisely this form. As we have seen, the dimensional reduction and simplified model can be understood as functions mapping from the feature space to the latent space, and from the latent space to the output space. Likewise, MBAM and the effective model can be understood as functions mapping from the original parameter space to the effective parameter space, and from the effective parameter space to the prediction space. The relations in figures \ref{fig:category} and \ref{fig:category2} seem to suggest a form of dependence where higher and lower level models retain an explicit functional relationship. 

Of course, in practice there \emph{is} a loss of information here: the process is not generally reversible, and so figures \ref{fig:category} and \ref{fig:category2} are themselves idealizations. After all, as we have seen, the key feature of these techniques is precisely such a reduction in the complexity of the models, and thus a loss of information. Renormalization group techniques explicitly reduce the information from the underlying theory, as we have seen in section \ref{sec:renormalizability}. Likewise, MBAM techniques reduce complexity by eliminating sloppy parameter combinations as we approach the manifold boundaries, retaining core predictive power in a lower-dimensional model without needing exact, fully derivable connections to microscopic details. These seem naturally understood as cases of approximate reduction, which can maintain essential dependencies across scales without requiring strict derivability. Dimensional reduction techniques also exhibit structural stability, capturing scale-invariant features in complex data. For example, in manifold learning lower-dimensional embeddings retain the essential topology and geometry of data. However, such idealizations are most naturally viewed as pragmatic simplification rather than barriers to reduction, which are made explicit in the functional forms relating the parameter spaces of the various models. As such, manifold learning seems highly amenable to  interpretation as a family of approximate Nagelian reductions (see \citealt{Dizadji-Bahmani2010} for a defense of this model of reduction). They seem to provide a family of test cases in which infinite idealizations do not present a natural barrier to reductive techniques. It would be an interesting line for future research to demonstrate this directly in particular examples, and to reconcile such a reductive approach with the idealizations involved.

Furthermore, the techniques discussed in this paper suggest that some of the philosophical debates around the renormalization group may apply more widely across other sciences. Recall that \citet{Wallace_naivete, PorterWilliams, Fraser_Realism, Fraser_RealProblem, Fraser_Towards, Miller} have argued that the success of effective field theories supports a form of selective scientific realism. My contention that effective theory construction can be understood as a special case of nonlinear dimensional reduction techniques suggests that such arguments can be applied more widely. It suggests that the success of effective theories is not unique to physics, but reflects a more general feature of successful scientific modeling: the ability to identify and preserve essential features while eliminating irrelevant degrees of freedom.

However, the machine learning perspective presented in this paper does not, in itself provide a defense against these kind of skeptical challenges provided by \citet{Ruetsche} and \citet{Sebastien_Rivat}.  For example, suppose that we wish to adopt a scientific realist perspective on some effective parameter combinations in our effective theory. We have a good reason for doing so: the theory leads to accurate predictions \textbf{and} we have reason to believe that this effective theory will remain a good effective model, even if our knowledge of the original model parameters changes. However, if our underlying theory changed more drastically, for example requiring entirely different parameters, there is no guarantee that this effective model will remain effective. In essence, effective theories remain vulnerable to \emph{unconceived alternative} theories, that lie outside of the model space under consideration (see \citealp{Freeborn-sloppy, Stanford_Exceeding}). 

If something akin to the global manifold hypothesis can be broadly defended, this might plausibly make room for a wider family of effective realist defenses of scientific theories, applying well beyond the scope of physics. After all, if the global manifold hypothesis holds, then many real-world datasets can be effectively compressed by dimensional reduction methods. As such, we might expect effective theory building techniques to be broadly successful precisely because many real-world target systems are amenable to them. Unconceived alternative theories might still replace our current best theories, but there would perhaps be less reason to expect them.  One might defend it with an inference to the best explanation: the manifold hypothesis is the best explanation for the remarkable success of manifold learning across a wide variety of domains. However, unfortunately, the global manifold hypothesis lacks a compelling theoretical motivation, with the main arguments being empirical \citet{Brahma2016, fefferman2016manifold,gorban2018manifold}.

Nonetheless, these techniques suggest one promising path for the selective realist in particular fields: to show that the salient epistemic features relevant to effective realism can also apply to a wider family of algorithmic reduction approaches, coupled with a suitable defense of a relevant local manifold hypothesis. Putting such an argument on a solid and rigorous footing would require substantial further work, but this suggests a potentially fruitful avenue for further research. There is not room to rigorously develop and defend such an argument here; however, \citet{Freeborn-sloppy} suggests one such possibility for extending effective realistic arguments beyond their traditional domain of quantum field theory.

\begin{appendices}

\end{appendices}

\bibliography{disbib}
\bibliographystyle{apalike}

\end{document}